\documentclass[twocolumn,showpacs,preprintnumbers,amsmath,amssymb]{revtex4-1}

\usepackage{amssymb}
\usepackage{amsmath}
\usepackage{graphicx}
\usepackage{color}
\usepackage{hyperref}
\usepackage{slashbox}
\usepackage[marginal]{footmisc}

\newcommand{\be}{\begin{equation}}
\newcommand{\ee}{\end{equation}}
\newcommand{\bea}{\begin{eqnarray}}
\newcommand{\eea}{\end{eqnarray}}
\newcommand{\beal}{\begin{aligned}}
\newcommand{\eeal}{\end{aligned}}

\usepackage{color}

\def\tr{\text{Tr}}

\begin{document}



\title{Unruh Quantum Otto heat engine with level degeneracy}


\author{Hao Xu}
\email{xuh5@sustech.edu.cn}
\affiliation{Institute for Quantum Science and Engineering, Department of Physics, Southern University of Science and Technology, Shenzhen 518055, China}
\affiliation{Department of Physics, University of Science and Technology of China, Hefei 230026, China}

\author{Man-Hong Yung}
\affiliation{Institute for Quantum Science and Engineering and Department of Physics, Southern University of Science and Technology, Shenzhen 518055, China}
\affiliation{Shenzhen Key Laboratory of Quantum Science and Engineering, Shenzhen, 518055, China}


\date{\today}

\begin{abstract}
We investigate the Unruh quantum Otto heat engine with level degeneracy. An effectively two-level system, where the ground state is non-degenerate and the excited state is $n$-fold degenerate, is acting as the working substance, and the vacuum of massless free scalar field serves as a thermal bath via the Unruh effect. We calculate the heat and work at each step of the Unruh quantum Otto cycle and study the features of the heat engine. The efficiency of the heat engine depends only on the excited energy values of the two-level system, not on its level degeneracy. However, the degeneracy acts as a kind of thermodynamic resource and helps us to extract more work than in the non-degenerate case. The extractable work has a finite upper bound, corresponding to $n\rightarrow \infty$.

\end{abstract}


\maketitle

\section{Introduction}

Quantum thermodynamics represents an overlapping area between two foundational physical theories developed independently, namely thermodynamics and quantum mechanics~\cite{Gemmer2004,Kosloff2013,Alicki2018}. Generally speaking, thermodynamics is a theory dealing with large-scale macroscopic processes, while quantum mechanics is a fundamental theory applicable to physical systems at the microscopic scale. The theory of quantum mechanics has achieved an unprecedented success in the history of physics. A fundamental question on the quantum theory is whether it is possible to derive all of the laws of thermodynamics from the established quantum principles, making quantum thermodynamics in the limelight \cite{Vedral2002,Brandao2013}.

In thermodynamics, one of the most important concept is the working mechanism of heat engines, which are systems converting heat or thermal energy into work or vice versa. For each heat engine, a working substance is required to absorb energy from a heat source, and generates work while transferring heat to a colder sink, returning to the initial state at the end. In classical thermodynamics, the heat engine can be described by a closed cycle in the state diagram. The Pressure-Volume ($P-V$) and the Temperature-Entropy ($T-S$) diagrams are the primary visualization tools for the study of heat engines. Carnot cycle, which consists of two isothermal steps and two adiabatic steps, is known to be an upper limit on the efficiency for any classical thermodynamic engine.

However, at the microscopic scale, the quantum mechanical laws should be taken into account for the working substances. Due to the recent advances in the technology of micro- and nano-machines, the study of quantum heat engine has attracted much attention \cite{Rezek2006,Quan2006,Quan2008,Abe2010,Thomas2010,Agarwal2013,Ro?nagel2013,Ivanchenko2014}. Particularly, we are interested in the following questions: whether the presence of quantum features, such as energy discreteness and level degeneracy, can affect the properties of thermodynamics for macroscopic objects? If so, how should the properties be modified? In fact, such kind of problems have been explored in 1950's \cite{Scovil1959}. Our work extends the results of Kieu, where a quantum heat engine was represented by a two-energy-eigenstate system (a qubit)~\cite{Kieu2005,Kieu2003}. There, a quantum Otto cycle was analyzed.

Recall that in classical thermodynamics, the Otto cycle consists of two adiabatic steps and two isochoric steps. In quantum thermodynamics, the quantum working substance is described by a density matrix $\rho(t)$ evolving under a time-dependent Hamiltonian ${\cal H}(t)$; the time evolution of the energy expectation value $ \langle E(t) \rangle = \tr (\rho(t){\cal H}(t))$ satisfies the following relation,
\begin{equation}
\partial_t \langle E(t) \rangle=\tr (\partial_t \rho(t){\cal H}(t)) + \tr (\rho(t)\partial_t{\cal H}(t)).
\end{equation}
The first term on the right-hand side is associated with the change of the population of $\rho(t)$, while the second term is with the Hamiltonian ${\cal H}(t)$.
We can identify the first one as the heat transfer $\langle Q \rangle$ and the second one as the work $\langle W \rangle$.
For the quantum Otto cycle, in each step either $\langle Q \rangle$ or $\langle W \rangle$ vanishes.

In contrast with classical Otto cycles, It was found~\cite{Kieu2005,Kieu2003} that for a qubit acting as the working substance, it is not sufficient to have the temperature $T_H$ of the heat source just hotter than temperature $T_C$ of the cold reservoir. There must be a minimum temperature difference between the $T_H$ and $T_C$, depending on the energy gaps in the qubit Hamiltonian spectrum before and after the adiabatic expansion, denoted by $\omega_1$ and $\omega_2$. Explicitly, in order to produce a finite amount of work, the condition $T_H>(\omega_2/\omega_1)T_C$ must be satisfied.

On the other hand, the field of quantum thermodynamics involves many fundamental questions, for example, the relationship between gravity theories and quantum field theory, which may shed some light on quantum gravity. In particular, applying the quantum field theory to curved space-time associated with a black hole, one finds that the quantum theory predicts the emission of particles by the black hole~\cite{Hawking:1974rv,Hawking:1974sw}. Similar analysis can also be applied to an accelerating observer in the Minkowski space-time, who observes thermal radiation from the vaccum, but an inertial observer would not. This is known as the Unruh effect \cite{Fulling:1972md,Unruh:1976db,Unruh:1977}. Even though the experimental test of the Unruh effect remains a challenge, models of the particle radiation detectors have been developed. In 1976, Unruh introduced a detector model consisting of a box containing a nonrelativistic particle satisfying the Schrodinger equation \cite{Unruh:1976db}. The Unruh effect can be detected if the particle jumps from the ground state to excited states. In 1979, DeWitt modified the model by replacing the particle with a two-level point monopole, thus Unruh effect is shown as the population change of the density matrix \cite{DeWitt1979}. Nowadays we generically call this model Unruh-DeWitt detector. See, e.g. ref. \cite{Felicetti:2015kta,Rodriguez-Laguna:2016kri,Wang:2014uoa,Ahluwalia:2015wha,Cozzella:2017ckb,Crispino:2007eb,Hu2019} for other proposals.

However, for a long time the study of connecting quantum thermodynamics and gravity theories remains little explored. This situation has changed until recent years. In \cite{Arias:2017kos} Arias et al. proposed that an Unruh-DeWitt detector undergoing acceleration can act as the working substance of a heat engine. An accelerated two-level system can extract work from the quantum field theory vacuum in a quantum thermodynamic Otto cycle. This model is known as Unruh quantum Otto heat engine. In this set up, the vacuum of the massless free field theory serves as a thermal bath via the Unruh effect. The two isochoric processes are achieved by linearly coupling the qubit undergoing higher and lower acceleration with the quantum vacuum, thus changing the population of the density matrix, while maintaining the energy level of the qubit, which implies vanishing work. The two adiabatic processes correspond to the increase and decrease of the excited energy level of the qubit, while keeping the density matrix unchanged, which implies vanishing heat. Moreover, Arias et al. derived the conditions to have both a thermodynamic and a kinematic cycle, which define a range of allowed accelerations for the Unruh engine. This model was extended to quadratically coupled scalar and fermionic fields in \cite{Gray:2018ifq}.

In the present work we consider the Unruh Quantum Otto heat engine with level degeneracy in the excited state. Within quantum thermodynamics, models with level degeneracy have been explored in thermal machines. It has been observed that this degeneracy may act as a kind of thermodynamic resource, helping us to extract more work than in the non-degenerate case \cite{Allahverdyan2007,Klimovsky2014,Mehta2017,Ma2017,Allahverdyan2013}. It is natural to consider an effectively two-level system such that while the ground state is non-degenerate, the excited state is $n$-fold degenerate, then we couple it linearly to a free massless scalar field prepared in its vacuum state. During the adiabatic steps, the two-level system undergoes an expansion/contraction of its excited level energy, while in the isochoric steps, the system is linearly coupled to the quantum vacuum with two different constant accelerations, making a closed Otto cycle, so that we can investigate the consequent modifications brought by the excited level degeneracy.

\section{Unruh Quantum Otto cycle}

In Unruh quantum Otto cycle, the working substance is an effectively point-like two-level system, where the ground state is non-degenerate and the excited state is $n$-fold degenerate. We denote the eigenlevels of system $\{|e_1 \rangle, \cdots, | e_n \rangle, | g \rangle \}$, which are associated with energies $\{ \omega, \cdots, \omega, 0 \}$.
Hence, the Hamiltonian of the system can be written as
\begin{equation}
{\cal H}=\omega \sum_{k=1}^{n} |e_k\rangle\langle e_k|.
\end{equation}
During the adiabatic steps, the energy $\omega$ changes, and we assume the quantum adiabatic theorem is valid in these processes. The initial density matrix $\rho_{in}$ for the system is given as
\begin{equation}
\rho_{in}=\sum_{k=1}^{n}\frac{p}{n} |e_k\rangle\langle e_k|+(1-p)|g\rangle\langle g|.
\end{equation}

The four steps of the Unruh quantum Otto cycle are
\begin{itemize}
\item \textit{Step} 1: The system moves with a constant velocity $v$ and density matrix $\rho_{in}=\sum_{k=1}^{n}\frac{p}{n} |e_k\rangle\langle e_k|+(1-p)|g\rangle\langle g|$. The excited energy value expands from $\omega_1$ to $\omega_2$. This is a adiabatic step and the density matrix is unchanged.	
\item \textit{Step} 2: The system is in contact with the quantum vacuum and undergoes a higher acceleration $\alpha_H$ from speed $v$ to $-v$. The final density matrix reads $\rho=\rho_{in}+\delta \rho_H$. This is a isochoric step and the excited energy value $\omega_2$ is unchanged.

\item \textit{Step} 3: The system is isolated from the quantum vacuum and moves with a constant velocity $-v$ and density matrix $\rho=\rho_{in}+\delta \rho_H$. The excited energy value reduces from $\omega_2$ to $\omega_1$. This is a adiabatic step and the density matrix is unchanged.	

\item \textit{Step} 4: The system is in contact with the quantum vacuum and undergoes a lower acceleration $\alpha_C$ from speed $-v$ to $v$. The final density matrix reads $\rho=\rho_{in}+\delta \rho_H+\delta \rho_C=\rho_{in}$. This is a isochoric step and the excited energy value $\omega_1$ is unchanged.
\end{itemize}

Here we give a brief discussion on the relativistic motion with constant acceleration $\alpha$. The space-time coordinates of the particle are
\begin{eqnarray}
t=\frac{1}{\alpha}\sinh(\alpha\tau),\qquad  x=\frac{1}{\alpha}\cosh(\alpha\tau),
\end{eqnarray}
where $\tau$ is the proper time. The velocity of the particle is $v=\tanh(\alpha\tau)$, so we can obtain the particle takes $2\tau=2 \operatorname{arctanh}(v)/\alpha$ time to change from velocity $-v$ to $v$ (or vice versa) with acceleration $\alpha$.

\subsection{Step 1}

In step 1, the density matrix is unchanged, thus there is no heat absorbed by the system. However, the energy expansion of the excited states from $\omega_1$ to $\omega_2$ is due to the cost of work, which is
\begin{equation}
\langle W_1\rangle=\tr (\rho_{in}\Delta{\cal H})=p\,\left(\omega_2-\omega_1\right).
\end{equation}

\subsection{Step 2}
In the step 2, we investigate the interaction between the two-level system and the free massless scalar field \cite{Birrell1982}. The field theory is prepared in its vacuum state. Due to the Unruh effect, when the two-level system moves with acceleration $\alpha$, the quantum vacuum will act as thermal states with temperature $T=\frac{\alpha}{2\pi}$, thus changing the population of the density matrix of the system. We denote the total Hamiltonian as
\begin{equation}
\mathbb{H}=\mathbb{H}_0+\mathbb{H}_{int},
\end{equation}
where the $\mathbb{H}_0$ is the sum of the two-level system Hamiltonian ${\cal H}$ and the free massless scalar field Hamiltonian ${\cal H}^{field}$, and $\mathbb{H}_{int}$ is the interaction Hamiltonian between the system and the field, which reads
\begin{equation}
\mathbb{H}_{int}= g\,m\varphi(\chi(\tau)).
\end{equation}
Here the $g$ is a weak coupling constant, $m$ is a monopole operator, and $\varphi(\chi(\tau))$ is the scalar field evaluated on the coordinate $\chi(\tau)=(t,x)$. The action of $m$ on the ground state $|g\rangle$ takes the system into a superposition of the $n$-fold degenerate excited states, with equal amplitudes $1/\sqrt{n}$, while the action on the excited states takes the system into the ground state $|g\rangle$. In matrix form, we obtain
\begin{equation}
m=\frac{1}{\sqrt{n}}\left(\begin{array}{cccc} {0} & {\cdots} & {0} & {1} \\
{\vdots} & {\ddots} & {\vdots} & {\vdots} \\ {0} & {\cdots} & {0} & {1} \\
{1} & {\cdots} & {1} & {0} \end{array}\right).
\end{equation}
If we denote $\varrho$ to be the density matrix of the total system, we have the quantum Liouville's equation in the interaction picture, which reads
\begin{equation}
i\frac{d\varrho_I(t)}{dt}=[\mathbb{H}_{int}^I,\varrho_I(t)],
\end{equation}
where $\varrho_I(t)=\mathbb{U}_0^{-1}(t,t_0)\varrho(t)\mathbb{U}_0(t,t_0)$ and $\mathbb{H}_{int}^I=\mathbb{U}_0^{-1}(t,t_0)\mathbb{H}_{int}\mathbb{U}_0(t,t_0)$. The initial density matrix
\begin{equation}
\varrho_I(t_0)=\varrho(t_0)=\varrho_{in}=\rho_{in}\otimes|0\rangle\langle0|.
\end{equation}
We expand the above Liouville's equation in terms of a Dyson series and apply the order product operator $T$, then the form of $\varrho_I(t)$ can be written as  \begin{eqnarray*}
\varrho_I(t)&&=\varrho_{in}-i\int_{t_0}^td\tau\left[\mathbb{H}_{int}^I(\tau),\varrho_{in}\right]\nonumber\\
&&-\frac{1}{2}\int_{t_0}^td\tau\int_{t_0}^t d\tau'T\left\{\left[\mathbb{H}_{int}^I(\tau),\left[\mathbb{H}_{int}^I(\tau'),\varrho_{in}\right]\right]\right\}+\ldots
\end{eqnarray*}
Since we focus on the population change of the two-level system, we can perform a partial trace over the field degrees of freedom, saying
\begin{equation}
\rho_I(t)=\tr_{\mathrm{field}}\,\varrho_I(t).
\end{equation}
We are applying a perturbative approach in small coupling constant $g$, so we obtain the final state up to second order. After some tedious calculations, we can have the final form of the density matrix reads
\begin{equation}
\rho_I(t)=\rho_{in}+\delta \rho_H(t)
\end{equation}
where
\begin{equation}
\delta \rho_H(t)=\delta p_H(t)\left(\begin{array}{cccc} {1/n} & {\cdots} & {1/n} & {0} \\
{\vdots} & {\ddots} & {\vdots} & {\vdots} \\ {1/n} & {\cdots} & {1/n} & {0} \\
{0} & {\cdots} & {0} & {-1} \end{array}\right).
\end{equation}
The exact form of $\delta p(t)$ is
\begin{equation}
\delta p_H(t)=g^2\int_{-\bar\tau}^td\tau\int_{-\bar\tau}^td\tau'\left((1-p)e^{-i\omega_2\Delta\tau}-\frac{p}{n}\,e^{i\omega_2\Delta\tau}\right)
G^+_{\alpha}(\tau,\tau'),
\label{ph}
\end{equation}
where $\Delta\tau=\tau-\tau'$ and $G^+_{\alpha}(\tau,\tau')$ is the two point correlation function for free massless scalar field.

Since this is isochoric step and the excited energy value $\omega_2$ is unchanged, there is no cost of work by the system. However, due to the population change $\delta \rho(t)$ of the density matrix, we have the heat absorbed by the system is
\begin{equation}
\langle Q_2\rangle=\tr {(\delta \rho_H(t)\cal H)}=\omega_2\delta p_H.
\end{equation}

\subsection{Step 3}
In the step 3, we isolate the system from the quantum vacuum, thus the density matrix $\rho=\rho_{in}+\delta \rho_H$ stays unchanged and there is no heat absorbed. We reduce the excited energy value from $\omega_2$ to $\omega_1$, so the work reads
\begin{equation}
\langle W_3\rangle=\tr ((\rho_{in}+\delta_H \rho(t))\Delta{\cal H})=(p+\delta p_H)\,\left(\omega_1-\omega_2\right).
\end{equation}

\subsection{Step 4}
In the step 4, we reconsider the interaction between the two-level system and the free massless scalar field, but now the acceleration is lower than the step 2 and the excited energy value is $\omega_1$. The method is the just same and we can obtain the value of $\delta p_C(t)$. It is worth noting that we cannot just replace the $p$ by $p+\delta p_H(t)$ in \eqref{ph}, since the initial density matrix $\rho_{in}+\delta \rho_H(t)$ for step 4 is not diagonal matrix. Further analysis shows that
\begin{eqnarray}
\delta p_C(t)&&=g^2\int_{-\bar\tau}^td\tau\int_{-\bar\tau}^td\tau'\Big((1-p-\delta p_H)e^{-i\omega_1\Delta\tau} \nonumber\\
&&-(\frac{p}{n}+\delta p_H)\,e^{i\omega_1\Delta\tau}\Big)G^+_{\alpha}(\tau,\tau').
\end{eqnarray}
However, since we are only considering the terms up to second order and $\delta p_H$ is already of order two, we can ignore this term safely. The $\delta p_C(t)$ reads
\begin{eqnarray}
\delta p_C(t)&&=g^2\int_{-\bar\tau}^td\tau\int_{-\bar\tau}^td\tau'\Big((1-p)e^{-i\omega_1\Delta\tau}\nonumber\\
&&-\frac{p}{n}\,e^{i\omega_1\Delta\tau}\Big)G^+_{\alpha}(\tau,\tau').
\end{eqnarray}
Although this formula is similar with $\delta p_H$ in step 2, in order to make sure the density matrix comes back to its initial state, we must have $\delta p_H+\delta p_C=0$. Since we set $\delta p_H>0$, we have $\delta p_C<0$. As before, there is no cost of work by the system, and the heat reads
\begin{equation}
\langle Q_4\rangle=\tr {(\delta \rho_C(t)\cal H)}=\omega_1\delta p_C.
\end{equation}

\section{Numerical results}

In the above section, we explicitly calculate the heat and work at each step of the cycle. Furthermore, in order to to obtain a closed
cycle, we require the density matrix comes back to its initial state in step 1, satisfying $\delta p_H+\delta p_C=0$. Since we set $\delta p_H>0$, we must have $\delta p_C=-\delta p_H<0$. The total mean heat absorbed by the two-level system reads
\begin{equation}
\langle Q\rangle=\langle Q_2\rangle+\langle Q_4\rangle=(\omega_2-\omega_1)\delta p_H.
\end{equation}
And the total work produced on the two-level system reads
\begin{equation}
\langle W\rangle=\langle W_1\rangle+\langle W_3\rangle=(\omega_1-\omega_2)\delta p_H.
\end{equation}
We can check that $\langle Q\rangle+\langle W\rangle=0$, which satisfies the first law of thermodynamics. The efficiency of the heat engine is defined as
\begin{equation}
\eta=\frac{\langle Q_2\rangle+\langle Q_4\rangle}{\langle Q_2\rangle}=1-\frac{\omega_1}{\omega_2}.
\end{equation}
This is our first main result, which indicates that \emph{the efficiency of the heat engine depends only on the excited energy value of the two-level system, not on its level degeneracy.}

However, this does not mean the exact value of $\delta p_H$ is insignificant, since $\delta p_H$ is proportional to the extractable work of the heat engine. As mentioned before, the $\delta p_H$ is written as
\begin{eqnarray}
\delta p_H(t)&&=g^2\int_{-\tau_H}^{\tau_H}d\tau\int_{-\tau_H}^{\tau_H}d\tau'\Big((1-p)e^{-i\omega_2\Delta\tau} \nonumber\\
&&-\frac{p}{n}\,e^{i\omega_2\Delta\tau}\Big)G^+_{\alpha_H}(\tau,\tau'),
\label{ph2}
\end{eqnarray}
where the interaction time
\begin{equation}
2\tau_H=2\operatorname{arctanh}(v)/\alpha_H,
\end{equation}
so that the two-level system undergoes the acceleration $\alpha_H$ from speed $v$ to $-v$. The method to calculate $\delta p_H(t)$ is discussed in the appendixes of \cite{Arias:2017kos,Gray:2018ifq}. Here we only present the basic steps and results. The finite-time integral from $-\tau_H$ to $\tau_H$ can be extended to infinite by introducing a regulator $\xi_{\tau_H}(\tau)$, which is a smooth function that is finite for $-\tau_H<\tau<\tau_H$, while being approximately zero outside the domain. Notice that the regulator must be a \emph{continuous} function, otherwise the excitation rate associated with accelerated finite-time detectors will be divergent \cite{Svaiter1992,Higuchi1993}. The behavior of the two level system does not depend sensitively on the particular choice of the regulator \cite{Higuchi1993}, thus we adapt the choice of \cite{Arias:2017kos}, choosing
\begin{equation}
\xi_{\tau_H}(\tau)=\frac{\tau_H^2}{\tau^2+\tau_H^2},
\end{equation}
then the $\delta p_H(t)$ becomes
\begin{eqnarray}
\delta p_H(t)&&=g^2\int_{-\infty}^{\infty}d\tau\int_{-\infty}^{\infty}d\tau' \xi_{\tau_H}(\tau)\xi_{\tau_H}(\tau') \Big((1-p)e^{-i\omega_2\Delta\tau}\nonumber\\
&&-\frac{p}{n}\,e^{i\omega_2\Delta\tau}\Big)G^+_{\alpha_H}(\tau,\tau').
\label{deltap}
\end{eqnarray}
Inserting the two point correlation function $G^+_{\alpha_H}(\tau,\tau')$ for the massless scalar field, which reads
\begin{eqnarray}
	G^+_{\alpha_H}(\tau,\tau')&&=\frac{1}{4\pi^{2}}\left(\frac{\alpha}{2i\sinh[(\alpha_H \Delta\tau-i\epsilon)/2]}\right)^{2}\;\nonumber\\
&&=-\frac{1}{4\pi^2}\sum_{k=-\infty}^{\infty}{\frac{1}{(\Delta\tau-i\epsilon-2\pi ik/\alpha_H)^2}},
\end{eqnarray}
into the $\delta p_H(t)$ and using the residue theorem, we can obtain
\begin{equation}
\delta p_H=g^2\left((1-p)J\left(-\frac{\omega_2}{\alpha_H},2\alpha_H\tau_H\right)-\frac{p}{n}\,J\left(\frac{\omega_2}{\alpha_H},2\alpha_H\tau_H\right)\right),
\label{ph3}
\end{equation}
where
\begin{eqnarray*}
J(x,y)&&=
\frac{(y/2)^2e^{-|x|y}}{16\sin^2(y/2)}+\frac{|x|y}{8}\theta(x)\nonumber\\
&&+
\frac{y^2e^{-2\pi|x|}}{64\pi^2}
\left(\phi(e^{-2\pi|x|},2,1+\frac{y}{2\pi})-\phi(e^{-2\pi|x|},2,1-\frac{y}{2\pi})\right)
\nonumber\\
&&+
\frac{|x|y^2e^{-2\pi|x|}}{32\pi}
\left(\phi(e^{-2\pi|x|},1,1+\frac{y}{2\pi})-\phi(e^{-2\pi|x|},1,1-\frac{y}{2\pi})\right),
\label{j}
\end{eqnarray*}
and
\begin{equation}
\phi(z,s,a)=\sum_{k=0}^{\infty}\frac{z^k}{(k+a)^s}
\end{equation}
is the Lerch-Hurwitz function. This result agrees with (C.16) of \cite{Gray:2018ifq}, but differs from \cite{Arias:2017kos} for a factor $1/2$, which is just a simple omission and does not affect the qualitative features of $\delta p_H$. There is also an extra $-1/8$ in \cite{Arias:2017kos}, which was introduced to resolve $\tau_H\rightarrow 0$ case. See. \cite{Gray:2018ifq,Sriramkumar:1994pb} for more discussion.

Since we have obtained the exact formula of $\delta p_H$, we can investigate the features of $\delta p_H$ and study its relations with other coefficients by using the numerical methods. The first thing we need to pay attention to is that $\delta p_H$ is obtained by the perturbation theory, thus we require $\delta p_H\ll p$. From the above formulas we can have $\delta p_H\sim g^2 \omega_2 \tau_H p=g^2\frac{\omega_2}{\alpha_H} \operatorname{arctanh}(v) p$. Defining $\frac{\omega_2}{\alpha_H}\equiv a_H$, we have
\begin{equation}
g^2a_H\operatorname{arctanh}(v)\ll 1.
\end{equation}
It is obvious that when the other coefficients are fixed, the larger $v$ brings larger interaction time $\tau_H$, so the absolute value $|\delta p_H|$ also grows larger. Without loss of generality, we set $v=0.99$, so that we can only consider the range of $0<a_H<0.05$, which is about one order of magnitude smaller than $1/g^2$. Now we are ready to study the relations between $\delta p_H$ and other coefficients.

\begin{figure}
\begin{center}
\includegraphics[width=0.45\textwidth]{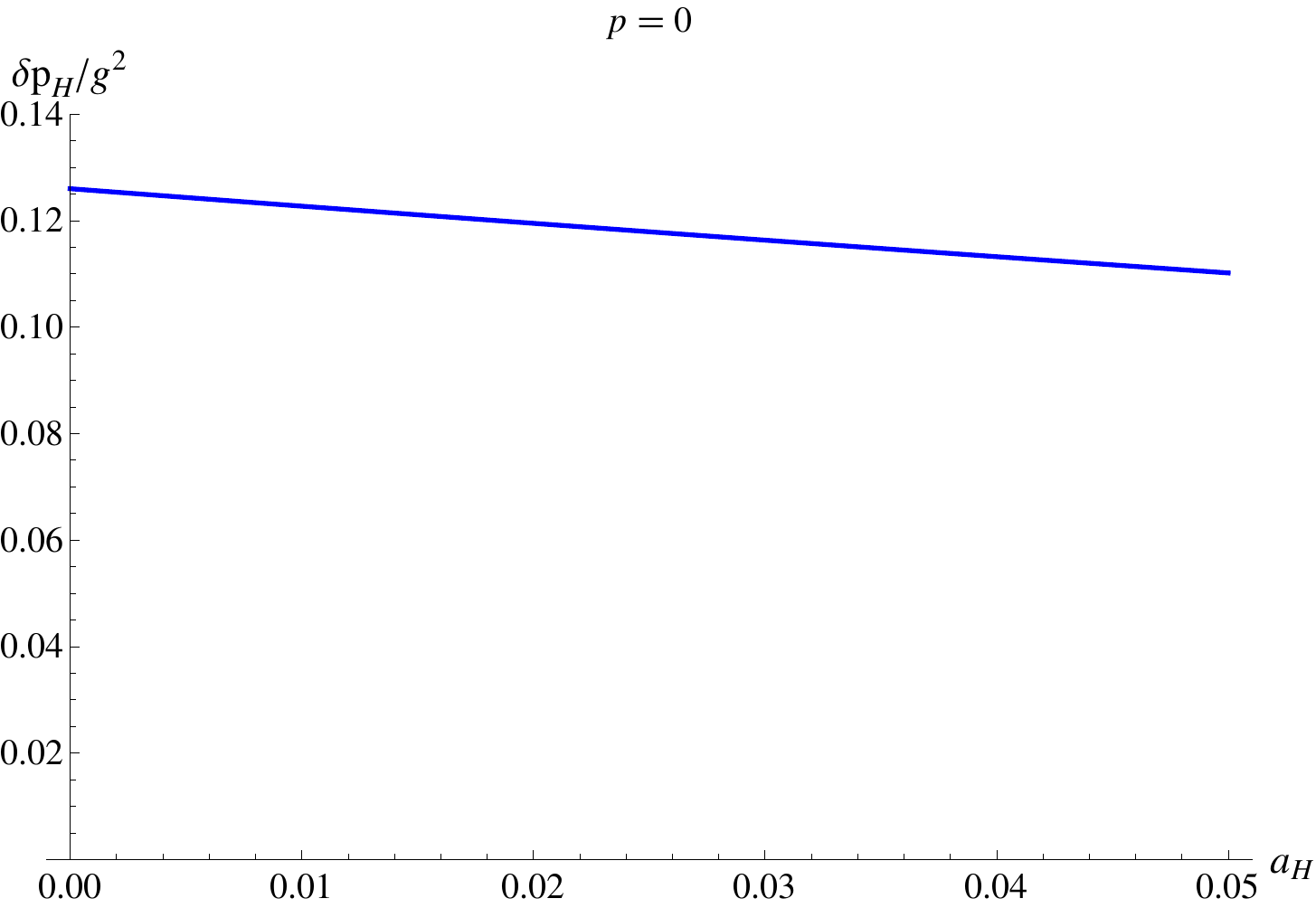}
\includegraphics[width=0.45\textwidth]{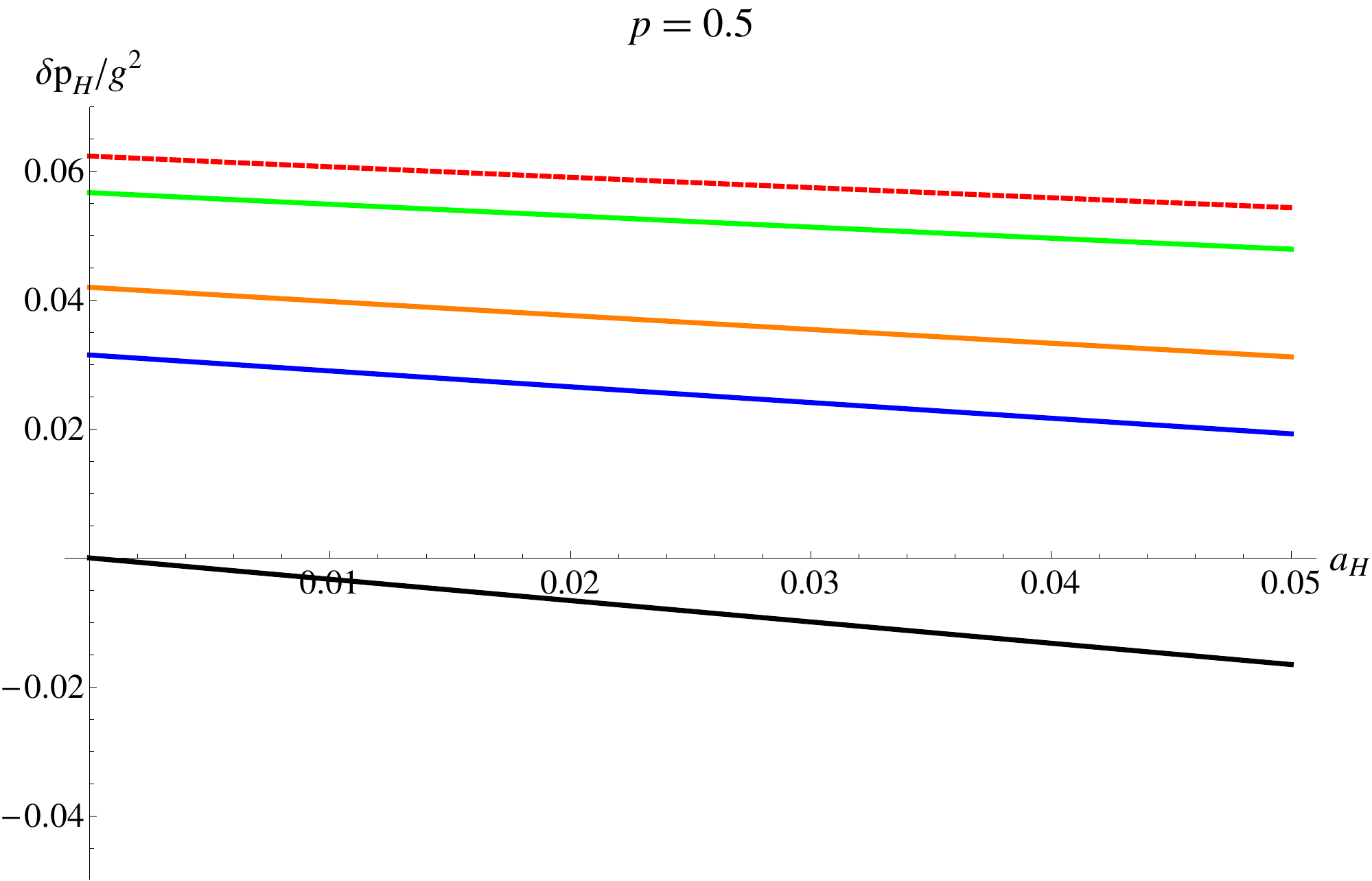}
\includegraphics[width=0.45\textwidth]{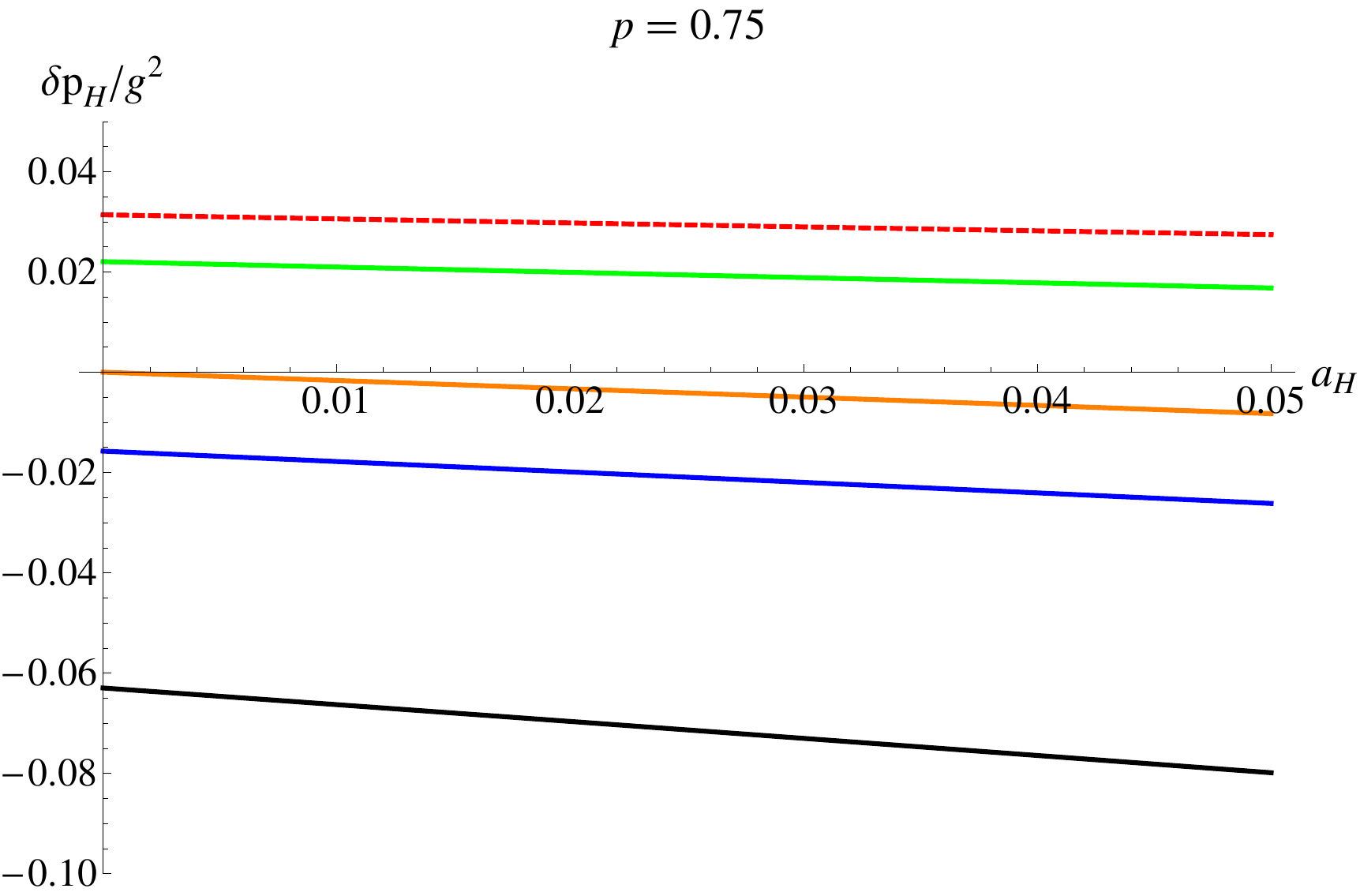}
\includegraphics[width=0.45\textwidth]{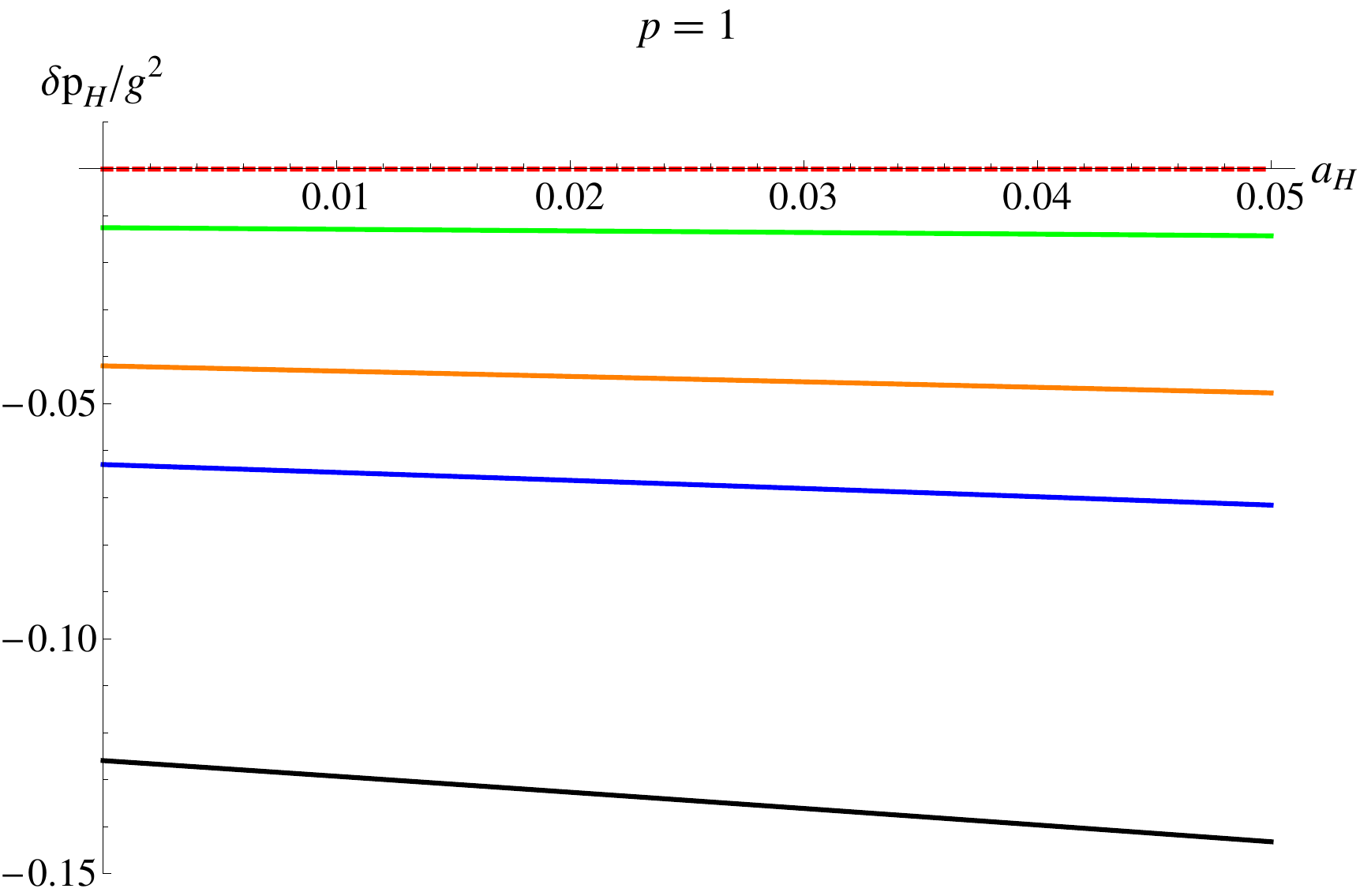}
\caption{Behavior of the $\delta p_H/g^2$ as function of $a_H\equiv \omega_2/\alpha_H$.}
\label{fig1}
\end{center}
\end{figure}

In Figure \ref{fig1} we present some examples on the behavior of the $\delta p_H/g^2$ as function of the $a_H\equiv \omega_2/\alpha_H$, corresponding to $p=0$, $p=0.5$, $p=0.75$ and $p=1$. For the case of $p=0$, the second term on the right-hand side of \eqref{ph3} vanishes, thus the degeneracy $n$ does not affect the value of $\delta p_H$. For any positive integer $n$, the formula of $\delta p_H$ stays the same, and it decreases when the $a_H\equiv \omega_2/\alpha_H$ grows, which
corresponds to the decreasing temperature, and thus agrees with the intuitive expectation. For the case of $p=0.5$, from bottom to top the solid lines correspond to $n=1$, $n=2$, $n=3$, $n=10$, and the dashed line corresponds to $n\rightarrow \infty$. As $a_H\rightarrow 0$, we can have the $\delta p_H/g^2$ reads
\begin{equation}
\delta p_H/g^2=0.126\big(1-p-\frac{p}{n}\big)
\label{a0}
\end{equation}
For $p=0.5$ and $n=1$, $\delta p_H/g^2=0$. When $a_H$ increases, $\delta p_H/g^2$ becomes negative, which indicates the two-level system can not absorb heat from the quantum field vacuum. When $n$ becomes larger, $\delta p_H/g^2$ is positive again, and it is bounded from above for the $n\rightarrow \infty$. Similarly, for the case of $p=0.75$, the $n=1$ and $n=2$ cases are always negative. For $n=3$ and $a=0$, $\delta p_H/g^2=0$. We can have positive $\delta p_H/g^2$ only for $n\geq 4$. For the case of $p=1$, the vacuum state is empty, thus we can never have the positive $\delta p_H/g^2$.

\begin{figure}
\begin{center}
\includegraphics[width=0.45\textwidth]{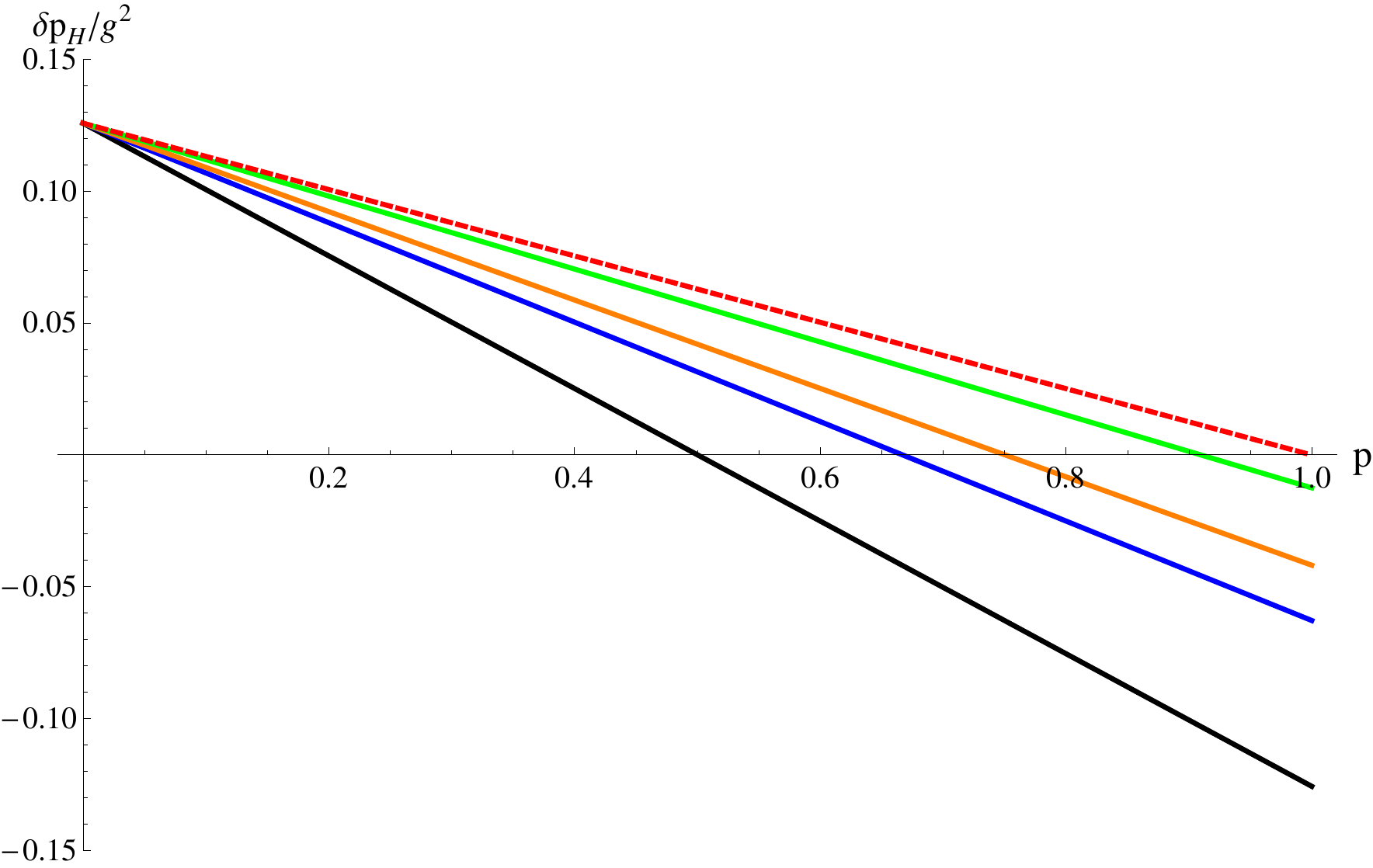}
\caption{Behavior of the $\delta p_H/g^2$ as function of $p$ when $a_H=0$.}
\label{fig2}
\end{center}
\end{figure}

In Figure \ref{fig2} we present the behavior of the $\delta p_H/g^2$ as function of $p$ when $a=0$, which corresponds to divergent temperature. From bottom to top the lines correspond to $n=1$, $n=2$, $n=3$, $n=10$ and $n\rightarrow\infty$. They all share the same value at $p=0$. When $p$ increases, the $\delta p_H/g^2$ decreases, and the lines with smaller $n$ always decrease faster. For the case of $n=1$, the $\delta p_H/g^2$ becomes negative when $p>0.5$. For general $n$, we know from \eqref{a0} that $\delta p_H/g^2$ remains positive if and only if
\begin{equation}
p<\frac{n}{n+1}.
\end{equation}
A larger $n$ ``dilutes'' the vacuum state in the Hilbert space, thus it can produce a larger amount of work and expands the range of $p$ which predicts the positive $\delta p_H/g^2$.

Similarly, we can also study the population change of the density matrix in step 4, where the acceleration $\alpha_C$ is lower and the excited energy $\omega_1$ is smaller. Using the same methods, we can have the $\delta p_C$ reads
\begin{equation}
\delta p_C=g^2\left((1-p)J\left(-\frac{\omega_1}{\alpha_C},2\alpha_C\tau_C\right)-\frac{p}{n}\,J\left(\frac{\omega_1}{\alpha_C},2\alpha_C\tau_C\right)\right).
\end{equation}
In order to make a closed Otto cycle, we must have
\begin{equation}
\delta p_H+\delta p_C=0,
\end{equation}
which serves as a constrained condition for the coefficients and it can be used to find the initial population of the excited state $p$. If we define
\begin{eqnarray}
{\cal P}\left(a_H,a_C,v\right)&&=\frac{4}{(a_H+a_C)\operatorname{arctanh}(v)}\Big(J\left(-a_H,2\operatorname{arctanh}(v)\right)\nonumber\\
&&+J\left(-a_C,2\operatorname{arctanh}(v)\right)\Big)
\end{eqnarray}
where $a_C\equiv \frac{\omega_1}{\alpha_C}$, then the $p$ reads
\begin{equation}
p=\frac{n{\cal P}}{(n+1){\cal P}+1}.
\end{equation}
We can also find that the $p$ must satisfy
\begin{equation}
p<\frac{n}{n+1}.
\end{equation}

\section{Summary}

In this work we investigate the Unruh quantum Otto heat engine with level degeneracy. We consider an effectively two-level system such that the ground state is non-degenerate, and the excited state is $n$-fold degenerate. This two-level system can act as the working substance in a quantum heat engine, and the vacuum of the massless free scalar field serves as a thermal bath via the Unruh effect. During the isochoric steps, the system is linearly coupled with the quantum vacuum and undergoes higher and lower acceleration, thus changing the population of the density matrix. During the adiabatic steps, the excited energy level is expanded and reduced, while keeping the density matrix unchanged. We explicitly calculate the heat and work at each step of the cycle and study the features of the engine. We can find the efficiency of the heat engine depends only on the excited energy value of the two-level system, not on its level degeneracy. However, when the $n$ is larger, the extractable work becomes larger, thus the degeneracy indeed act as a thermodynamic resource and help us to extract more work than in the non-degenerate case. This extractable work has a finite bound, corresponding to $n\rightarrow \infty$. The larger $n$ also expands the range of the coefficient that can predict the positive work, which means by utilizing the level degeneracy, the value of $\delta p_H$ can become positive in cases where it would have been negative. Using a detector with a degenerate excited state, it is possible to extract work in scenarios which otherwise would require an input of work.

It is worth emphasizing that in the present work we adapt the smooth switching function, which is used to expand the finite-time integral to infinity, so that we can apply the Cauchy-Riemann residue theorem to evaluate the integral. When the interacting time goes to infinity, the switching function equals $1$ and the question reduces to the calculating of response function \cite{Birrell1982}
\begin{eqnarray}
\int_{-\infty}^{\infty}d\tau\int_{-\infty}^{\infty}d\tau'e^{-i\omega(\tau-\tau')}G^+_{\alpha_H}(\tau,\tau').
\end{eqnarray}
However, though the switching function decays quickly outside the region of interest, they are not of compact support. Strictly speaking, the switching function provides an approximation of the finite-time interaction in the framework of infinite-time interaction. Unfortunately, since our analysis is based on the infinite integral and Cauchy-Riemann residue theorem, closing the contour in an infinite semi-circle in the half of the complex $\Delta \tau$ plane, the detailed investigation on the contributions of subregion on $\tau$ is beyond our reach. So far the best we can do is, assuming the switching function $\xi_{\tau_H}(\tau)\equiv\xi\left(\frac{\tau}{\tau_H}\right)$ satisfies $\xi(0)=1$ and $\xi'(0)=0$, to expand the switching function as a Taylor series around $\tau\approx 0$, i.e.
\begin{eqnarray}
\xi\left(\frac{\tau}{\tau_H}\right)&&\approx\xi(0)+\xi'(0)\left(\frac{\tau}{\tau_H}\right)+\frac{1}{2}\xi''(0)\left(\frac{\tau}{\tau_H}\right)^2+O\left(\frac{\tau^n}{\tau_H^n}\right)\nonumber\\
&&=1+\frac{1}{2}\xi''(0)\left(\frac{\tau}{\tau_H}\right)^2+O\left(\frac{\tau^n}{\tau_H^n}\right).
\end{eqnarray}
Inserting the above formula into eq.\eqref{deltap}, we can calculate perturbatively the corrections from the switching function and study the difference between using the switching function and the infinite-time interaction. This issue has been studied in \cite{Sriramkumar:1994pb}, though the detailed calculation can be quite troublesome. The questions of the switching function are very interesting and we will consider them in our future work.

Similar analysis may also be extended to other kinds of heat engines with various working substance, quantum field theories, kinematic and thermodynamic cycles. It would be interesting to consider such examples and to study their features. Since in our model we find the degeneracy helps us to extract more work because larger $n$ ``dilutes'' the vacuum state in the Hilbert space, we believe additional energy levels have the similar effect. For example, if the system has $k$ non-degenerate excited eigenstates and it reaches equilibrium with heat source, the population on the excited states will be $\frac{\sum^k_{i=1}e^{-\omega_i/T}}{1+\sum^k_{i=1}e^{-\omega_i/T}}$. Obviously a larger $k$ also allows more population on the excited states, and the features of heat transfer depend on the specific initial state and the interaction between working substance and the quantum field theory. Working substance located in different backgrounds, such as curved space-time with black holes, can also be investigated.  These researches may shed more light on the relations between quantum thermodynamics and gravity theories. In this work we use the perturbative method and the interaction Hamiltonian must be turned on and off during the isochoric steps. The issue of the energy costs associated with performing the cycle would be interesting \cite{Kerstjens2017}. We hope to be able to consider these questions in our future work.

\begin{acknowledgments}
We would like to thank the referee whose suggestions and comments helped us in improving the original manuscript. Hao Xu thanks Yuan Sun and Yuan Sheng Wang for useful discussions. We acknowledge the support by the National Natural Science Foundation of China (No.11875160), the NSFC Guangdong Joint Fund (U1801661), Natural Science Foundation of Guangdong Province (2017B030308003), the Guangdong Innovative and Entrepreneurial Research Team Program (No.2016ZT06D348), and the Science Technology and Innovation Commission of Shenzhen Municipality (ZDSYS20170303165926217, JCYJ20170412152620376, JCYJ20170817105046702).
\end{acknowledgments}

\providecommand{\href}[2]{#2}\begingroup\raggedright\endgroup

\end{document}